\documentclass{elsart}

\usepackage{natbib}

\usepackage{epsfig}

\usepackage{amssymb}
\def\dj{\hbox{d\kern-0,347em \vrule width0,3em height1,252ex
depth-1,21ex \kern0,051em}}
\begin{document}

\begin{frontmatter}



\title{Crack avalanches in the three dimensional random fuse model}


\author[1]{Stefano Zapperi}
\author[2]{Phani Kumar V.V. Nukala}
\author[2]{Sr{\dj}an \v{S}imunovi\'{c}}
\address[1]{INFM UdR Roma 1 and SMC, Dipartimento di Fisica,
Universit\`a "La Sapienza", P.le A. Moro 2, 00185 Roma, Italy}
\address[2]{Computer Science and Mathematics Division, 
Oak Ridge National Laboratory, Oak Ridge, TN 37831-6359, USA}

\begin{abstract}
We analyze the scaling of avalanche precursors
in the three dimensional random fuse model by numerical simulations. 
We find that both the integrated and non-integrated avalanche size
distributions are in good agreement with the results of the
global load sharing fiber bundle model, which represents the mean-field
limit of the model.
\end{abstract}

\begin{keyword}
fracture \sep random fuse model \sep avalanches
\PACS 46.50.+a \sep  64.60.Ak
\end{keyword}

\end{frontmatter}

% main text
\section{Introduction}

Understanding the scaling properties of fracture in disordered media
represents an intriguing theoretical problem with some technological
implications \cite{breakdown}. Of particular interest is the acoustic
emission (AE) recorded in a stressed material before failure. The
noise is a consequence of micro-cracks forming and propagating in the
material and provides an indirect measure of the damage accumulated in
the system. For this reason, AE is often used as a non-destructive
tool in material testing and evaluation.  The distribution of crackle
amplitudes follows a power law, suggesting an interpretation in terms
of scaling theories.  This behavior has been observed in several
materials such as wood \cite{ciliberto}, cellular glass
\cite{strauven}, concrete \cite{ae} and paper \cite{paper}.

The statistical properties of fracture in disordered media 
are captured qualitatively  by lattice models,  describing the medium as 
a discrete set of elastic bonds with randomly 
distributed failure thresholds \cite{breakdown,fuse,duxbury}. 
In the simplest approximation of a scalar displacement,
one recovers the random fuse model (RFM) where a lattice of fuses with
random thresholds are subject to an increasing external current
\cite{fuse,duxbury}. Fracture of the RFM is preceded by avalanches of failure  
events \cite{hansen,zrvs,alava} which are reminiscent of the acoustic emission
activity observed in experiments. The distribution of avalanche sizes (i.e. the number
of bonds participating in an avalanche)  follows a power law. Initially two 
dimensional simulations yielded an exponent close to $\tau=5/2$ \cite{zrvs}, 
the value expected in the fiber bundle model (FBM) \cite{dfbm,hansen1}. 
In that model load is redistributed equally to all the 
fibers, representing thus a sort of mean-field limit of the RFM \cite{zrvs}. 
More recent large scale simulations, however, displayed significant (non-universal)
deviations from the mean-field result \cite{cond-mat}. Only some preliminary results are reported
in the literature for three dimensions 
\cite{rai-98}. Here we show that avalanches in the three dimensional
RFM follow quite closely the mean-field predictions. This is partly expected
since normally scaling exponents tend to the mean-field limit as the 
lattice dimensionality increases, although the exact value for the
upper critical dimension is not known for this problem.

\section{The random fuse model}

In the random thresholds fuse model  \cite{fuse,duxbury}, 
the lattice is initially fully intact
with bonds having the same conductance, but the bond breaking
thresholds, $t$, are randomly distributed based on a thresholds
probability distribution, $p(t)$.  The burning of a fuse occurs
irreversibly, whenever the electrical current in the fuse exceeds the
breaking threshold current value, $t$, of the fuse. Periodic boundary
conditions are imposed in both of the horizontal directions to simulate an
infinite system and a constant voltage difference, $V$, is applied
between the top and the bottom of the lattice system bus bars.

Numerically, a unit voltage difference, $V = 1$, is set between the
bus bars and the Kirchhoff equations are solved to determine the
current flowing in each of the fuses. Subsequently, for each fuse $j$,
the ratio between the current $i_j$ and the breaking threshold $t_j$
is evaluated, and the bond $j_c$ having the largest value,
$\mbox{max}_j \frac{i_j}{t_j}$, is irreversibly removed (burnt).  The
current is redistributed instantaneously after a fuse is burnt
implying that the current relaxation in the lattice system is much
faster than the breaking of a fuse.  Each time a fuse is burnt, it is
necessary to re-calculate the current redistribution in the lattice to
determine the subsequent breaking of a bond.  The process of breaking
of a bond, one at a time, is repeated until the lattice system falls
apart. In this work, we assume that the bond breaking thresholds are 
distributed based on a uniform probability distribution, 
which is constant between 0 and 1.

Numerical simulation of fracture using large fuse networks is often
hampered due to the high computational cost associated with solving a
new large set of linear equations every time a new lattice bond is
broken. Although the sparse direct solvers presented in \cite{nukalajpamg} are 
superior to iterative solvers in two-dimensional lattice systems, for 3D lattice 
systems, the memory demands brought about by the amount of fill-in during the 
sparse Cholesky factorization favor iterative solvers. Hence, iterative solvers are 
in common use for large scale 3D lattice simulations. The authors have developed 
an algorithm based on a block-circulant preconditioned 
conjugate gradient (CG) iterative scheme \cite{nukalajpamg2} for simulating 
3D random fuse networks. The block-circulant preconditioner was shown to be 
superior compared with the {\it optimal} point-circulant preconditioner for 
simulating 3D random fuse networks \cite{nukalajpamg2}. 
Since the block-circulant and {\it optimal} point-circulant preconditioners 
achieve favorable clustering of eigenvalues (in general, the more clustered 
the eigenvalues are, the faster the convergence rate is),  in comparison with 
the Fourier accelerated iterative schemes used for modeling lattice breakdown
\cite{bat-98}, this algorithm significantly reduced the
computational time required for solving large lattice systems. 

Using the algorithm presented in \cite{nukalajpamg2}, we have performed
numerical simulations on 3D cube lattice
networks.  For many lattice system sizes,
the number of sample configurations, $N_{config}$, used are
extremely large to reduce the statistical error in the numerical
results. In particular, we used $N_{config}=40000,3840,512,128,32$ for
$L=10,16,24,32,48$ respectively.

\section{Avalanches}

\begin{figure}[hbtp]
\centerline{\psfig{file=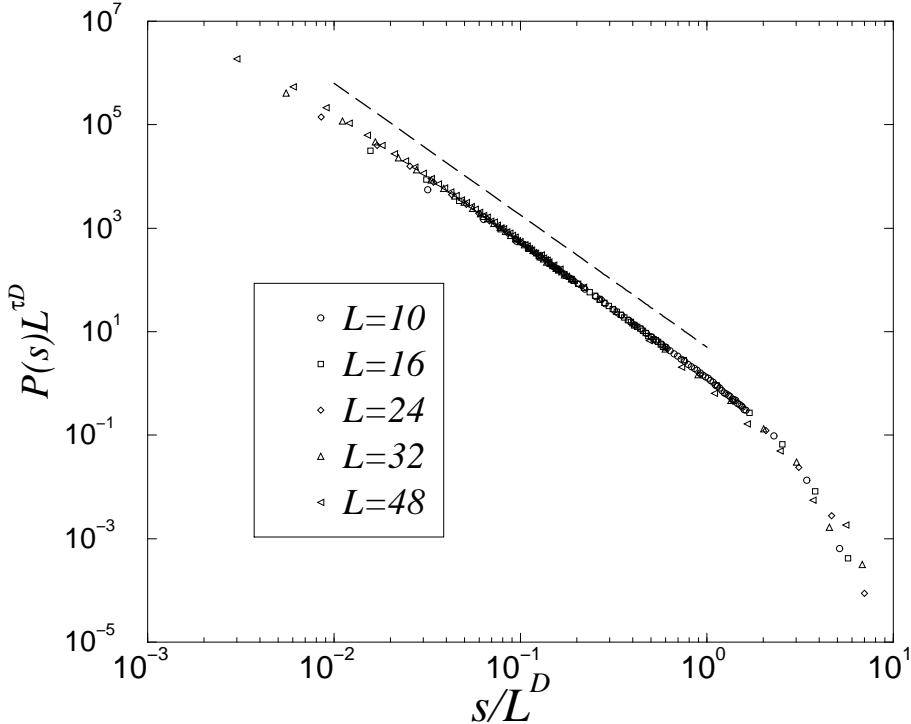,width=12cm,clip=!}}
\caption{Data collapse of the integrated avalanche size distributions. The exponents used for
the collapse are $\tau=2.5$ and $D=1.5$. The line with a slope $\tau=2.55$
is the best fit for the power law decay.}
\label{fig:1}
\end{figure}

When the current is increased at an infinitesimal rate
failure events cluster in the form of avalanches.
The typical avalanche size increases with the current up to the last catastrophic
failure event.  The avalanche size distribution is a power
law followed by an exponential cutoff at large sizes. The cutoff size $s_0$ 
is increasing with the lattice size, so that we can describe the distribution
by a scaling form 
\begin{equation}
P(s,L)=s^{-\tau} g(s/L^D),
\end{equation}
where $D$ represents the fractal dimension of the avalanches. 
To confirm this finite size scaling assumption, 
we perform a data collapse imposing the mean-field exponent $\tau=5/2$ 
and choosing $D=1.5$ (see Fig.~\ref{fig:1}). A direct power law fit of
the distribution yields instead $\tau=2.55$.

\begin{figure}[hbtp]
\centerline{\psfig{file=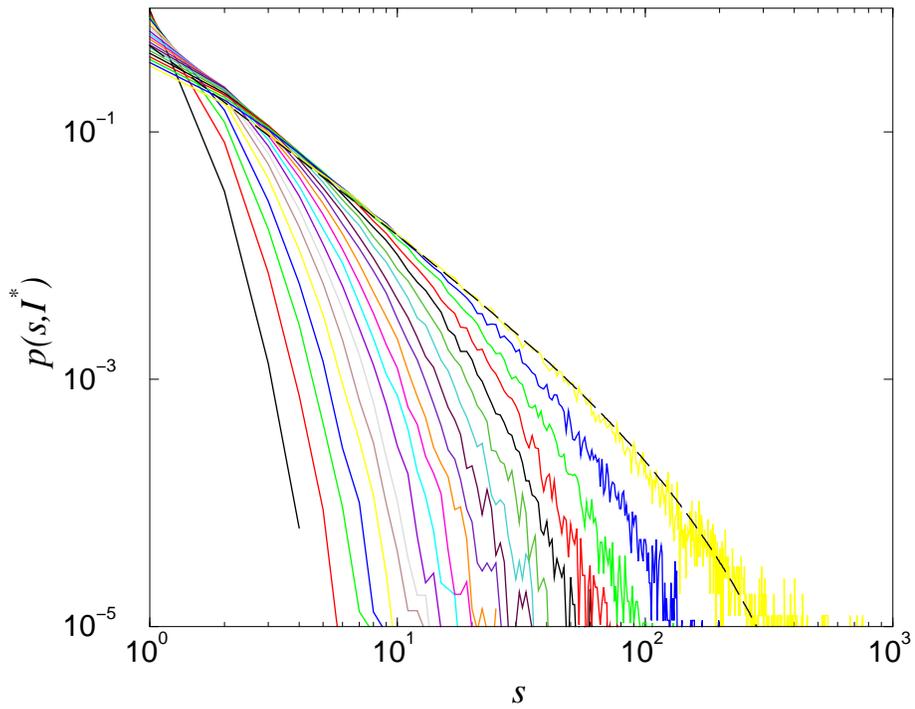,width=12cm,clip=!}}
\caption{The avalanche size distributions sampled over a small bin of the reduced current
$I^*$ for a cube lattice of size $L=48$. The dashed line is a fit according to
Eq.~\protect\ref{eq:binsize} with $\gamma=3/2$.
 }
\label{fig:2}
\end{figure}

We have considered avalanche statistics integrating the distribution
over all the values of the current, but the avalanche signal
is not stationary: as the current increases so does the avalanche size. 
In Fig.~\ref{fig:2} we report the distribution of avalanche sizes sampled at different
values of the current $I$. For each sample, we normalize the current by its maximum value $I_c$
and divide the $I^*=I/I_c$ axis in 20 bins. We then compute the avalanche size distribution
$p(s,I^*)$ for each bin and average over different realization of the disorder.
 The distribution follows a law of the type
\begin{equation}
p(s,I^*)= s^{-\gamma}\exp(-s/s^*),\label{eq:binsize}
\end{equation}
with $\gamma \simeq 1.5$ and $s^*$ is an increasing function of $I^*$, 
in good agreement with mean-field results.

\section{Conclusions}

We have performed numerical simulations of the random fuse model in three
dimensions, focusing on the avalanche distributions. The scaling of the
distributions is well captured by mean-field theory. This is in contrast
with the behavior in two dimensions that shows larger deviations \cite{cond-mat}. This can
be expected on general grounds since typically the mean-field limit is 
approached as the dimensions are increased.

% \appendix

% \section{}
% \label{}

% Bibliographic references with the natbib package:
% Parenthetical: \citep{Bai92} produces (Bailyn 1992).
% Textual: \citet{Bai95} produces Bailyn et al. (1995).
% An affix and part of a reference:
%   \citep[e.g.][Ch. 2]{Bar76}
%   produces (e.g. Barnes et al. 1976, Ch. 2).

{\bf Acknowledgment} PKVVN and SS are sponsored by the
Mathematical, Information and Computational Sciences Division, Office
of Advanced Scientific Computing Research, U.S. Department of Energy
under contract number DE-AC05-00OR22725 with UT-Battelle, LLC.

\end{document}